\documentclass[pre,floatfix,twocolumn,showpacs,nofootinbib]{revtex4-1}
\usepackage{amsmath}
\usepackage{amsfonts}
\usepackage{mathrsfs}
\usepackage{epsfig}
\usepackage{graphicx}
\usepackage{color}

\begin{document}

\title{Path lengths, correlations, and centrality in temporal networks}

\author{Raj Kumar Pan}
\author{Jari Saram\"aki}
\affiliation{%
BECS, School of Science and Technology, Aalto University, P.O.  Box 12200, FI-00076
}%
\date{\today}

\begin{abstract}
  In temporal networks, where nodes interact via sequences of
 temporary
events, information or resources can only flow through paths that follow
the time-ordering of events. Such temporal paths play a crucial
role in dynamic processes. However, since networks have so far
been usually  considered static or quasi-static, the properties of temporal paths 
are not yet well understood. 
Building on a definition and algorithmic
implementation of the average temporal distance between nodes, we study
temporal paths in empirical networks of human communication and air
transport. Although temporal distances correlate with static graph
distances, there is a large spread, and nodes that appear close from the
static network view may be connected via slow paths or not at all. Differences between static
and temporal properties are
further highlighted in studies of the temporal closeness centrality.
In addition, correlations and heterogeneities in the underlying event sequences
affect temporal path lengths, increasing temporal distances in
communication networks and decreasing them in the air transport network. 
\end{abstract}


\maketitle

\section{Introduction}
Understanding complex networks is of fundamental importance for studying
the behavior of various biological, social and technological
systems~\cite{Newman06a,Dorogovtsev03,Newman10}.  Often, networks represent
the complex lattices on which some dynamical processes unfold~\cite{BBV08},
from information flow to epidemic spreading. For such processes,
networks have mainly been considered static or quasi-static, such that
dynamic changes of the network structure take place at a time scale longer
than that of the studied process, and thus a node may interact with any or all of 
its neighbors at any point in time. 
In empirical analysis of systems where time-stamped data is available, a common 
approach has been to integrate connections or interaction events over the period of observation. This
results in a static network where a pair of nodes is connected by a link
if an event has been observed between them at any point in time.
The frequency of events between nodes
may then be taken into account with link weights that represent the number of events
between nodes (see, e.g.,~\cite{Barrat04,Onnela07})
Taking a step beyond static networks, in the dynamic network view (see, e.g.~\cite{Gautreau09,Rosvall10}), links
are allowed to form and terminate in time, such as friendships forming 
and decaying in social networks. This view is commonly adopted in 
epidemiological modeling in the form of \emph{concurrency} or \emph{transmission} graphs~\cite{Morris95,Riolo01}
-- e.g. for sexually transmitted diseases, links represent partnerships
that have a beginning and an end, and the prevalence of multiple
simultaneous partnerships has significant effects on the dynamics
of outbreaks.

However, there are many cases where even the dynamic network picture
is too coarse-grained, as the nodes are in reality connected 
 by recurrent, temporary \emph{events} of short duration at specific times only~\cite{Kempe00,Kempe02,Holme05a,Vazquez07,Iribarren09,Karsai10,HolmePNAS10, HolmeSimulated10}. 
We use the term \emph{temporal network} for such
systems to distinguish them from static or (quasi-static) dynamic networks.
The events in a temporal network represent the temporal sequence of interactions
between nodes, and thus the dynamics of any process mediated by such interactions
depends on their structure. As an example,  in an 
air transport network, events may represent individual flights transporting passengers
In a social network, events may represent individual social 
interactions (phone calls, emails, physical proximity) that 
allow information to propagate through the network from one individual to another.
In epidemiological modeling, data on the timings of possible transmission events, \emph{i.e.} individual encounters that
may result in disease transmission, has allowed for moving beyond the concurrency
graph view~\cite{HolmePNAS10, HolmeSimulated10}.




An immediate consequence of event-mediated interactions for any dynamics is
that it has to follow time-ordered, causal paths~\cite{Kempe02,Holme05a}.
Because of the causality requirement, the static network representation
where nodes are connected if any interaction has been observed between them
at any point in time can be misleading: although node $i$ may be connected
to node $j$ via some path in the static network, that path may not exist in
its temporal counterpart. Nevertheless, were the interaction events
uncorrelated and uniformly spread in time, they could in many cases be
taken into account by assigning weights to the edges of the static network,
so that the weights would represent the frequencies of events between
nodes~\cite{Barrat04,Onnela07} and regulate the rate of interactions.
However, it has turned out that this is commonly not the case: it has been
observed that for the dynamics of spreading of computer viruses,
information, or diseases, timings of the actual
events and their temporal
heterogeneities~\cite{Moody02,HolmePNAS10,HolmeSimulated10,Vazquez07, Iribarren09, Karsai10,Miritello10} play an
important role -- e.g., the burstiness of human communication has been
observed to slow down the maximal rate of information spreading~\cite{Iribarren09, Karsai10,Miritello10}.  Hence,
for a detailed understanding of such processes, one should adopt the
temporal network view. 

A temporal network can be represented by a set of $N$ nodes between which a
complete trace of all interaction events $\mathcal E$ occurring within the
time interval $[0,T]$ is known. Each such event can be represented by a
quadruplet $e \equiv (u,v,t,\delta t)$, where the event connecting nodes
$u$ and $v$ begins at $t$ and the interaction is completed in time $\delta
t$. As an example, $\delta t$ may correspond to the duration of a flight in
an air transport network or the time between an user sending an email and
the recipient reading it.  Broadly, we define $\delta t$ such that if an
event $e$ transmits something from $u$ to $v$, the recipient receives the
transmission only after a time $\delta t$. However, in some cases, events
can be approximated as instantaneous so that $\delta t = 0$ and they can be
represented with triplets $e \equiv (u,v,t)$, as in Ref.~\cite{Holme05a}. 
Further, events can be directed or undirected depending on whether the
transmission or flow is directed or not.

In some earlier papers~\cite{Tang10a,Tang10,Kostakos09}, temporal networks
have been represented as a set of graphs $\mathcal{G} = \langle G_0, \dots,
G_T \rangle$, where $G_t = (V_t,E_t)$ is the graph of pairwise interactions
between the nodes at time $t \in [0, T]$. Here, $V_t$ and $E_t$ represent
the nodes and edges at time $t$, respectively.  However, this picture is
only meaningful when the events are instantaneous (and, for practical
purposes, only when the time is discretized). If the events have a duration
$\delta t$ such a representation cannot be applied: it is not compatible
with the fact that for anything to be transmitted via node $i$ to node $j$,
$i$ has to receive the transmission before the event connecting $i$ and $j$
is initiated, but $j$ then receives the transmission only after a time
$\delta t$.

In this paper, we set out to study the time-ordered paths that span a
temporal graph and their durations.  Any dynamical processes have to
proceed along such paths;  consider, as an example, the deterministic
susceptible-infectious (SI) dynamics, where infected nodes always infect
their susceptible neighbors as soon as they interact. The speed of such
dynamics depends on how long it on average takes to complete time-ordered
shortest paths between nodes, \emph{i.e.} the average \emph{temporal
distance} between nodes, which in turn depends on the temporal
heterogeneity and correlations of the event sequence. As an example, 
in a social network, where events such as calls or emails mediate information, the
average temporal distance measures the shortest time it takes for any information
to be passed from one individual to another, either directly or via intermediaries. 
For other dynamics,
additional constraints can be placed on allowed transmission paths: e.g.,
for the susceptible-infectious-recovered (SIR) spreading dynamics where an
infected node remains infectious for a limited period of time only, there is
a waiting time threshold between consecutive events spanning a path.

We begin by defining the average temporal distance between nodes that
properly takes the finiteness of the period of observation into account. We
also present an algorithm for calculating such distances in event
sequences, based on the concept of vector clocks.  We then compare static
and temporal distances in empirical networks of human communication and air
transport and illustrate the differences.  We next turn to the role of
heterogeneities and correlations in the event sequences, and show that
their effects are strikingly different in our empirical networks: contrary
to the known effect of correlations slowing down dynamics in human
communications, they give rise to faster dynamics in the air transport
network.  The roles of correlations are also studied on temporal paths
constrained by a SIR-like condition on allowed waiting times between
events. Finally, we study the temporal centrality of nodes, and show that
nodes that may appear insignificant from the static point of view may in
fact provide fast temporal paths to all other nodes.

\section{Measuring distances in temporal graphs}
\label{sec:methods}

\subsection{Temporal paths and temporal distances: definitions}


Information or resources can be transmitted from node $i$ to node $j$ in a
temporal network only if they are joined by a  causal temporal path,
\emph{i.e.} a time-ordered sequence of events beginning at $i$ and ending
at $j$~\cite{Kempe02,Holme05a}.  If the events are non-instantaneous, a
temporal path exists only if there is a time-ordered sequence where each
event begins only after the previous one is completed\footnote{This requirement comes from our view of an event
as the ``fundamental unit" of interaction -- an email user may forward information obtained from an email only
after she has received and read it, and a passenger may only board a connecting flight if the 
previous flight arrives before the connecting flight departs. On the contrary,  \emph{e.g.}~in concurrency graphs where
a link in essence represents a string of interactions, it would
make sense to allow paths via temporally overlapping links.}. As an example,
suppose that there is an event $e_1 = (i, j, t_1, \delta t_1)$ between
nodes $i$ and $j$ and another event $e_2 = (j, k, t_2, \delta t_2)$,
between $j$ and $k$. This sequence of events spans the temporal path
$i\rightarrow j \rightarrow k$ only if $t_2 > t_1+\delta t_1$, and the time
it takes to complete this path, \emph{i.e.}~the temporal path length, is
then $\Delta t = t_2-t_1+\delta t_2$.  Let us define the \emph{temporal
distance} $\tau_{ij}(t)$ between $i$ and $j$ as the shortest time it takes
to reach $j$ from $i$ at time $t$ along temporal
paths\footnote{ Note that temporal
distances are inherently non-symmetric and generally, $\tau(i,j) \neq \tau(j,i)$. 
Thus the temporal distance defined here is not, strictly speaking,
a metric, and we use the term distance similarly to the geodesic graph distance
in directed networks.}.
If the fastest
sequence of events, \emph{i.e.}~the shortest temporal path joining $i$ and
$j$ begins at time $t'>t$ and its duration is $\delta t$, then
$\tau_{ij}=\left(t'-t\right)+\delta t$.  It is evident that this temporal
distance depends on the time of measurement $t$; it may also happen that no
such path exists and then $\tau_{ij}(t)=\infty$.  As $\tau_{ij} (t)$ is not
constant in time, it is useful to characterize temporal distances with an
\emph{average temporal distance} $\tau_{ij}$, averaged over the entire
period of observation.  However, taking this average is not straightforward
and certain choices have to be made.

For empirical event sequences, the period of observation $[0,T]$ is 
always finite\footnote{Evidently, the length of the period should be chosen such
that enough events are collected for any measure to be meaningful.
This problem is equally important for static network analysis, although it
is typically neglected, and made more difficult by the fact that there
may be changes in the system dynamics on multiple, overlapping time scales.
Here, we adopt the view that the defined measures are
estimates based on the events observed within a period of length $T$ and their
values are with certainty only representative for this window, although certain 
probability distributions be stationary across time. This is the approach typically taken
in studies of static networks aggregated over time, although it is seldom
explicitly stated.}. Because of this, the total number of
future events decreases as time increases, and consequently so does the
likelihood of the existence of a time-ordered path between any pair of
nodes. Thus, infinite temporal distances $\tau_{ij} (t)=\infty$ become increasingly
common when $t$ approaches $T$. There are three possible ways of taking
these infinite distances into account: (i) for each pair of nodes, averaging 
only over the range where $\tau_{ij}(t)$ is finite, as was done in Ref.~\cite{Holme05a},
(ii) getting rid of all infinite distances by 
assuming that the entire event sequence may be periodically repeated, \emph{i.e.}
assuming network-wide periodic temporal boundary conditions,
and (iii) handling the finite window size and infinite distances \emph{separately} for each 
pair of nodes $i$ and $j$ for which $\tau_{ij}$ is calculated, by assuming that
the observation window provides a good estimate of the frequency and duration of
paths for each node pair. 

\begin{figure}
\begin{center}
  \includegraphics[width=0.85\linewidth]{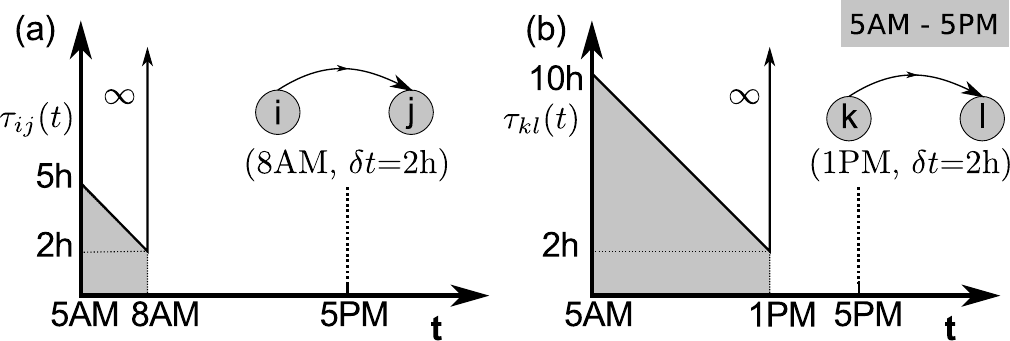}
\end{center}
\caption{Schematic representation of the variation of temporal distances
between two pairs of nodes, (a) $i$-$j$ and (b) $k$-$l$. The period of
observation is between 5AM and 5PM. In panel (a), the two nodes are
connected by an event that begins at 8AM and takes two hours to completion.
In panel (b), the nodes are connected by an event of the same duration at
1PM. If the average temporal distance were defined only over its finite
range, then $\tau_{ij}<\tau_{jk}$, although both pairs are connected via
similar events.}
\label{fig:schematic_tdist1}
\end{figure}

Let us first take a look at option (i), averaging the temporal distance only over 
the period where it is finite. The problem with this approach is that
 it introduces a bias in favor of temporal paths taking place early within
the period of observation.  This can be illustrated with a simple example
(see Fig.\ref{fig:schematic_tdist1}): suppose that node $i$ directly
interacts with $j$ only once at $t_1$, nodes $k$ and $l$ interact once at
$t_2$, and no other temporal paths exist between these nodes. Here,
$\tau_{ij}$ equals the shaded area divided by $t_1$ $(t_2)$. Now, if $t_1
\ll t_2$, the above averaging would imply that $\tau_{ij} \ll \tau_{kl}$,
because when the distances are finite, $\tau_{ij}(t) \ll \tau_{kl}(t)
\forall~t$. 

On the basis of the above, we now set the following requirement for the average temporal distance
$\tau_{ij}$: for any sequence of shortest temporal paths, the resulting
average temporal distance should not depend on when that sequence takes
place within the period of observation.  Hence, $\tau_{ij}$ should be the
same for both cases in Fig.~\ref{fig:schematic_tdist1}. This leaves
 us with options (ii) and (iii).  
 Both choices fulfill the above criterion for the simple example of Fig.~\ref{fig:schematic_tdist1}. However,
option (ii) can be ruled out by the following requirement:
nodes that are not connected via a temporal path within the observation
window should not become connected by applying the condition. If
the entire event sequence is periodically repeated, this is not the case,
as disconnected nodes may become connected via paths that may even
span multiple window lengths. Thus, in order to avoid unnecessary artifacts to the extent that is possible,
we base our definition of the average temporal distance on option (iii),
where the finite period of observation is handled 
separately for each pair of nodes. Specifically, for calculating $\tau_{ij}$,
we assume that  if there is a temporal path between $i$ and $j$ which begins 
at $t=t_1$ and the period of observation is $[0,T]$, then this temporal path will reoccur at time
$t=T+t_1$ \emph{without} affecting the paths or distances between any other pair of nodes. It is
easy to see that for the simple example of Fig.~\ref{fig:schematic_tdist1}, this is analogous to
assuming that we have a correct estimate of the frequency and duration of temporal paths
between $i$ and $j$.

\begin{figure}
\begin{center}
  \includegraphics[width=0.65\linewidth]{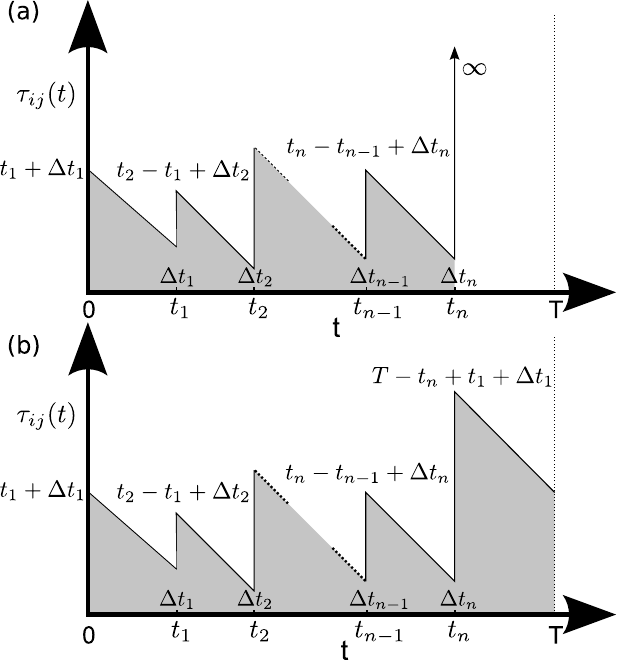}
\end{center}
\caption{Schematic representation of the time variation of the temporal
distance between a pair of nodes, $i$-$j$: (a) the actual distance and (b)
the distance with periodic boundary condition on paths connecting $i$ and
$j$.}
\label{fig:schematic_tdist2}
\end{figure}

Let us next have a closer look at how $\tau_{ij} (t)$ varies with time $t$
(see Fig.~\ref{fig:schematic_tdist2}) in a setting where there are several
shortest temporal paths at different points in time. Suppose that there is a temporal
path along a time-ordered sequence of events starting at time $t_1$ through
which one can reach node $j$ from $i$.  If the time of completion of this
path is $\Delta t_1$, then $\tau_{ij} (t_1) = \Delta t_1$. If this is the
only temporal path between $i$ and $j$ within the observation period, then
for any $t<t_1$, $\tau_{ij}(t) = (t_1-t) + \Delta t_1$, and for any
$t>t_1$, $\tau_{ij}(t)=\infty$. In general, if there are multiple shortest
temporal paths between nodes $i$ and $j$ that begin at times $t_1,\dots,t_n$
and have durations $\Delta t_1, \dots, \Delta t_n$, respectively, then the
temporal distance curve has the shape depicted in
Fig.~\ref{fig:schematic_tdist2}~(a). Application of the node-pair-specific boundary condition, \emph{i.e.}
repeating the first path, makes the temporal distance
between nodes $i$ and $j$ behave as depicted in
Fig.~\ref{fig:schematic_tdist2}~(b)\footnote{Note that periodic boundary conditions on the entire
event sequence, i.e.~repeating the sequence, could change the behavior near $T$, as
entirely new temporal paths that cross the boundary might appear.}. If there are $n$ shortest temporal
paths between $i$ and $j$ within the observation period, with beginning
times $t_1, \dots, t_n$ and durations $\Delta t_1, \ldots, \Delta t_n$,
then the average temporal distance is given by
\begin{equation}\label{eq:avg_t}
  \begin{split}
  \tau_{ij} &=  \frac{1}{T} \left[ t_1 \left(\frac{t_1}{2}+\Delta t_1
  \right) + \left(t_2-t_1\right) 
  \left(\frac{t_2-t_1}{2} + \Delta t_2\right) \right. \\ 
  &+ \dots + (t_n - t_{n-1})\left(\frac{t_n - t_{n-1}}{2} + \Delta
  t_n\right) \\ 
  &+ \left. (T-t_n) \left(\frac{T-t_n}{2} + t_1 + \Delta t_1 \right) \right].
  \end{split}
\end{equation}
If there is only one temporal path between these nodes, the above equation
reduces to $\tau_{ij} = \frac{T}{2} + \Delta t$, which is independent of
the actual time of occurrence of the path, fulfilling the criterion that average
temporal distance should be independent of the placement of the event
sequence within the observation window. 

\begin{figure*}
\begin{center}
  \includegraphics[width=0.30\linewidth]{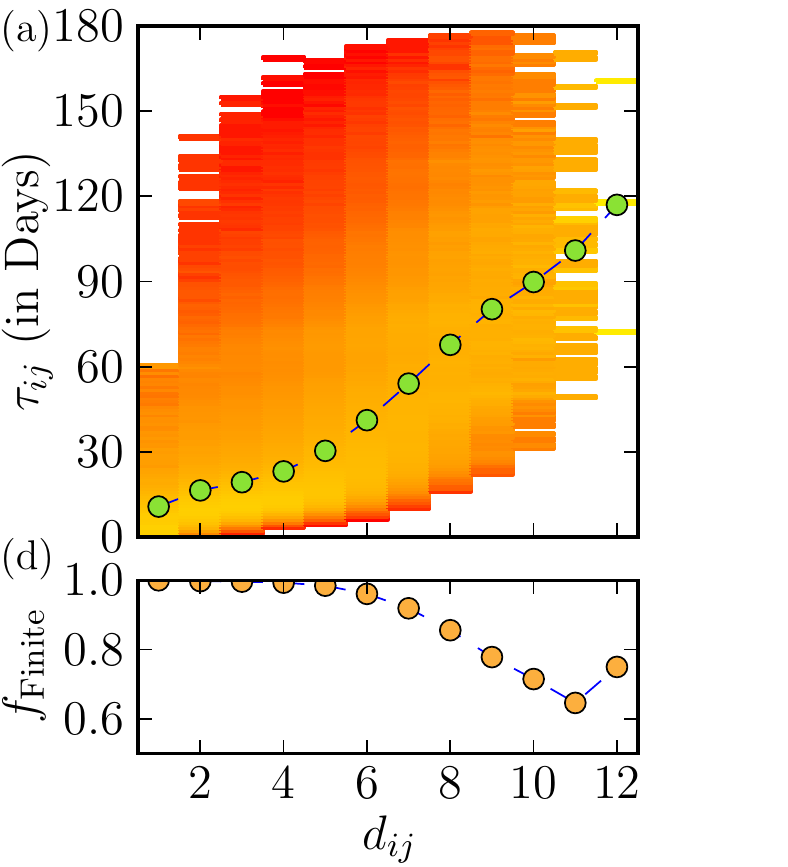}
  \includegraphics[width=0.3\linewidth]{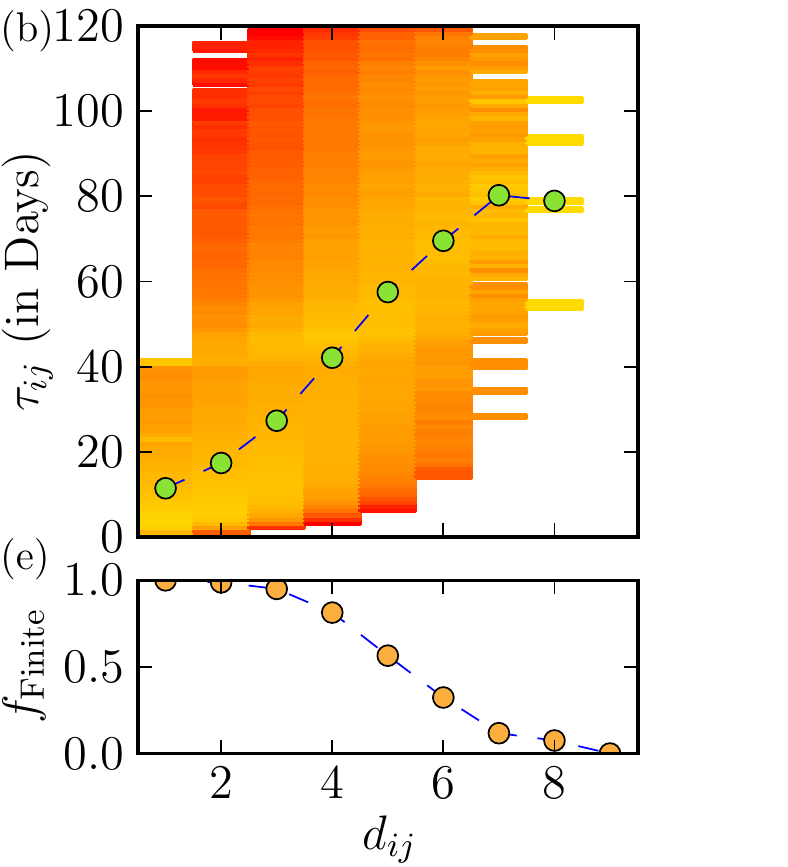}
  \includegraphics[width=0.3\linewidth]{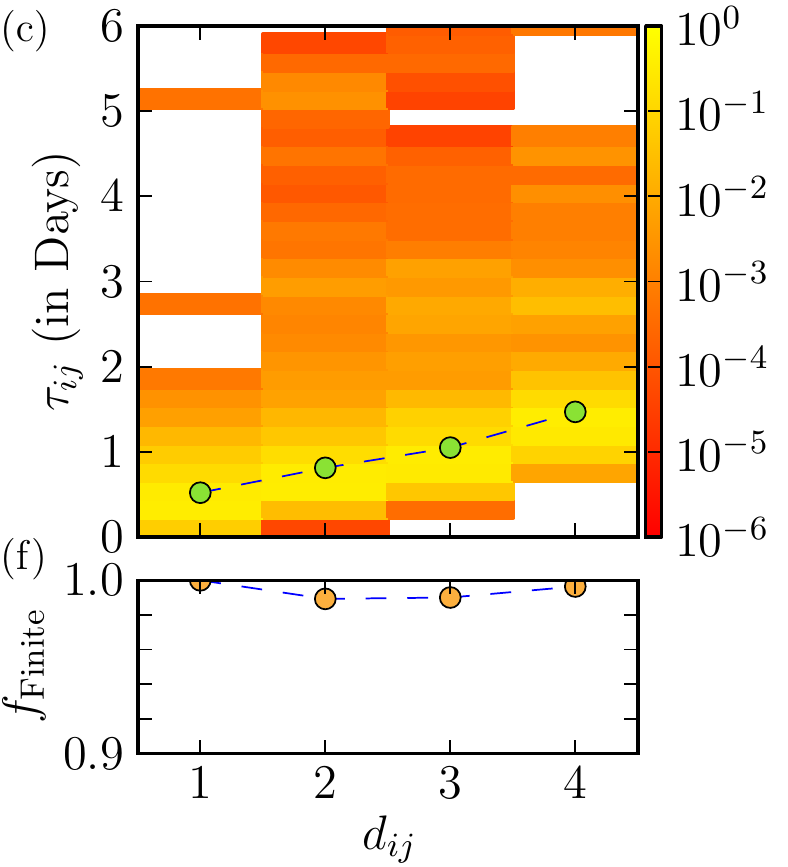}
\end{center}
\caption{Top: the average temporal distance, $\tau_{ij}$, against the
static graph distance, $d_{ij}$, between all pair of nodes for (a) the
call network, (b) the e-mail network, and (c) the air transport network.
The average temporal distances $\tau_{ij}$ were calculated using periodic
boundary conditions, as detailed in the text. The colors represent the
conditional probabilities of $\tau_{ij}$ for a given $d_{ij}$. Note the
broad distribution of $P\left(\tau_{ij}|d_{ij}\right)$ in all three cases.
Bottom: The fraction of finite temporal paths, $f_{\mathrm{Finite}}$, as a
function of $d_{ij}$ for (d) the mobile phone call network, (e) the e-mail
network and (f) the air transport network. It is seen that the longer a
static path, the less likely the existence of a corresponding temporal
path within the observation window.}
\label{fig:tdist}
\end{figure*}

\subsection{An algorithm for calculating temporal distances}

For calculating the above-defined average temporal distance between any two
nodes $i$ and $j$ in an empirical event sequence, we need to detect the beginning times of all
shortest temporal paths between $i$ and $j$ (i.e., $t_1, \dots, t_n$, and
the corresponding temporal distances at that particular time (i.e.,
$\tau_{ij}(t_1)=\Delta t_1,\dots,\tau_{ij}(t_n)=\Delta t_n$). Here, we use
the notion of vector clocks~\cite{Mattern89} and propose an algorithm 
for efficient calculation of these quantities. For describing the algorithm,
we use the metaphor of events transmitting information between nodes.

Let us assign a vector ${\pmb{\phi_{i}}}$ for each node, such that its element
$\phi_{i}^{j}(t)$ denotes the nearest point in time $t'>t$ at which node $j$
can receive information transmitted from node $i$ at time $t$, either
via a direct event or a time-ordered path spanned by any number of events.
We also define $\phi_{i}^{i}(t)=t$.
We then take advantage of a simple and efficient algorithm~\cite{Lamport78,
Mattern89,Kossinets08} to compute the shortest temporal paths between all
nodes within a finite time period $[0,T]$.  This is done by sorting the
event list in the order of decreasing time (i.e. ``backwards'') and going
through the entire list of events once.  Initially, we set all elements
$\phi_i^j=\infty \forall i\neq j$ at $T$, indicating that no node can
obtain any information, even indirectly, from any other after the end of
our observation period $T$.  Let us first assume that all events are
instantaneous and undirected, i.e., information flows in both directions.
We now go through the time-reversed event list event by event. For each
event $(i,j,t)$ we compare the vector clocks of $i$ and $j$ element-wise,
i.e., $\phi_{i}^{k}$ and $\phi_{j}^{k}$ $\forall k$, and update both with
the lowest value. If $\phi_{i}^{k}$ is updated, this indicates that the
event has given rise to a new shortest temporal path between $i$ and $k$
that begins at time $t$, and the associated temporal distance
$\tau_{ik}(t)=\phi_{i}^{k}(t)-t$. As the event connects $i$ and $j$, we
also set $\phi_{i}^{j}(t)=\phi_{j}^{i}(t)=t$, and thus
$\tau_{ij}(t)=\tau_{ji}(t)=0$. As each update of the vector indicates the
existence of a new temporal path, the updates define the beginning times
$t_1,\ldots, t_n$ and durations $\Delta t_1, \ldots, \Delta t_n$ of
temporal paths in the sum of Eq.~\ref{eq:avg_t}, allowing for computing the
average temporal distance between $i$ and $j$.

The algorithm can also be generalized for directed events with specific
durations. For details, see Appendix A. 

\section{Temporal paths and distances in empirical networks}

\subsection{Data description}
\label{sec:data}
In the following, we apply the above measures in the analysis of empirical
data on temporal graphs. We have chosen two very different types of data
sets: social networks, where information spreads through communication
events in time, and an air transport network, where events transport
passengers between airports.  For each data set, we consider the respective
temporal graph, \emph{i.e.}~the sequence of events, as well as its
aggregated static counterpart where nodes are linked if an event joining
them is observed in the sequence at any point in time.

Our first data set consists of time-stamped mobile phone call
data over a period of 120 days~\cite{Karsai10}, where each event corresponds to a voice call between
two mobile phone users. We consider the events here as undirected and
instantaneous, such that events may immediately transmit information. Note
that although calls have in reality a duration, one person participates in one call only at
a time, and thus for temporal paths, this duration can be
neglected.  For this study, we have selected a group of 1982 users that
comprise the largest connected component (LCC) of an aggregated undirected
network of users with a chosen zip code. Between these 1982 users, there
are 5420 undirected edges, containing in total 153045 calls. This network
is mutualized, \emph{i.e.} we retain only events associated with links
where there is at least one call both ways.  Our second social network data
set is an email network constructed from time-stamped email records of
university users~\cite{Eckmann04} within a period of 81 days. We consider
emails as directed and study only the Largest Weakly Connected Component
(LWCC) of the aggregated network, retaining events between its members,
arriving at 2993 users connected by 28843 directed edges with 202687
emails. Third, we consider an air transport network, where the flights
between all the airports in the US~\cite{BTS} for a period of 10 days
between 14th and 23rd December 2008 are observed. The air transport network comprises 279
airports connected via 4152 directed edges and altogether 180192 flights;
although edges are directed, 99.5\% of them are reciprocated. In the static
network, all airports belong to the Strongly Connected Component (SCC). All
times are converted to GMT.

We note that for the two social networks, the observation periods (120 and 81 days)
have been determined by the availability of data: we have chosen to use all
the data available to us. For the air transport network, because of the 
inherent periodicity of flight schedules, a shorter window was chosen.

\subsection{Relationship between temporal and static distances}
\label{sec:comparison}

Let us first consider the relationship between static and average temporal
distances in the empirical systems, $d_{ij}$ in the aggregated network and
$\tau_{ij}$  in the temporal graph (Fig.~\ref{fig:tdist} a,b,c).  Here, the
static distance is defined as usual as the number of links along the
shortest path connecting nodes in the aggregated network.
For the call and email data sets, 
the average temporal distance can be considered
as a measure of the time it takes for information to reach one node from another,
if it is transmitted via calls or email such that recipients pass on the information. 
For the air transport network, the average temporal distance measures the average
time to reach one airport from another, either directly or via connecting flights.
In all cases, the static distance measures the number of links one has to traverse
to get from one node to another. For a pair of
nodes joined by such a path, the shortest temporal paths may of course
follow another sequence of links, or not exist at all.  One would still
expect that in general, nodes that are far from each other in the static
network would also have large temporal distances.  For all three networks,
we find that on average this is indeed  the case  (Fig.~\ref{fig:tdist}
a,b,c) -- however, as the conditional distributions $P(\tau_{ij}|d_{ij})$
clearly show,  there is surprisingly large variation around the average in
all cases. As an example, in the mobile call network, there are node pairs
that are at the same graph distance $d_{ij}$, but whose temporal distances
differ by a factor of $10^2$. Likewise, one can find node pairs with a
relatively short temporal distance that are either directly linked or 10
links apart in the static network. This highlights the importance of the
temporal graph approach for processes whose dynamics depend on event
sequences: e.g., for any spreading process on such systems, the pathways
taken and the structure of the resulting branching tree can be entirely
different if shortest temporal paths are followed.

For the social networks, the relationship between the static and average
temporal distances is not linear, as there is an apparent increase in the
slope for larger temporal distances. Furthermore, the fraction of node
pairs at a given static distance that are also connected via a temporal
path, $f_{\mathrm{Finite}}$, is seen to decrease for higher static
distances (Fig.~\ref{fig:tdist} d,e,f). Hence, in social communication
networks, information between node pairs at large static distances may be
on average transmitted only slowly or not at all. However, for the mobile
call network, 95\% of node pairs are nevertheless connected via a temporal
path as very large static distances are infrequent; for the directed email
network, the corresponding fraction is lower, 58\%.
Note that the behavior of $f_{\mathrm{Finite}}$ depends on the
length of the observation period (120 days for the call network and 81 days for
the email network), and, in general, the frequency of events.  In addition, 
for the email network, the number of existing paths is naturally constrained by the 
directedness of the events, as from the point of view of information spreading,
emails carry the information one way only, whereas calls may transfer information
both ways. Thus, in the mobile call network, information may in theory be passed
from almost any node to any other within the period of observation, whereas
in the email network studied here this is not the case. Nevertheless, for both
systems, an observation window spanning several months does not guarantee
that all nodes are connected by a temporal path.
 On the contrary,
reflecting its function and design, in the air transport network almost all
pairs of nodes at any static distance are joined by a temporal path within
the 10-day period of observation.
\subsection{Effects of correlations on temporal distances}
\label{sec:dynamics}
The empirical event sequences in our datasets that span the temporal paths
contain correlations and heterogeneities affecting the temporal distances.
First, events follow  strong daily patterns. In the mobile call network,
the call frequency shows a peak around lunch time and early evening
(see~\cite{Karsai10}), whereas the frequency of flights is almost constant
during the day.  In the night, calls and departures of flights are
infrequent. Second, in addition to the daily pattern, there are other
non-uniformities in the event sequence: especially in human communications,
\emph{bursty} behavior giving rise to broad distributions of inter-event
times is common~\cite{Barabasi10, WuPNAS2010, Karsai10}.  Third, there are
event-event correlations, where one event may trigger another one, or
events have been scheduled such that one follows another. Such correlations
give rise to short waiting times between consecutive events along temporal
paths.

The effects of heterogeneities and correlations on temporal distances can
be investigated by applying null models where the original event sequences
are randomized to systematically remove these
correlations~\cite{Holme05a,Karsai10}.  Here, we apply null models that
separately destroy the following correlations: bursty or periodic event
dynamics on single links, event-event correlations between links and, and
the daily patterns. All structural properties of the static network are
retained, as the null models only modify the times of events between nodes.
The null models are as follows: (i) In the \emph{equal-weight link-sequence
shuffled} model, whole single-link event sequences are randomly exchanged
between links having the same number of events. Event-event correlations
between links are destroyed. (ii) In the \emph{time-shuffled} model, the
time stamps of the whole event sequence are shuffled. In this case, the
bursts, periodicity and the event-event correlations are destroyed, while
the daily patterns are retained. (iii) In the \emph{random-time} model, the
time stamps of all the events are chosen uniformly randomly from the period
of observation. Here, all temporal correlations including the daily cycle
are destroyed. When the events have a duration $\delta t$, this value
remains attached to each event whenever the time of its occurrence changes.

\begin{figure}
\begin{center}
  \includegraphics[width=0.95\linewidth]{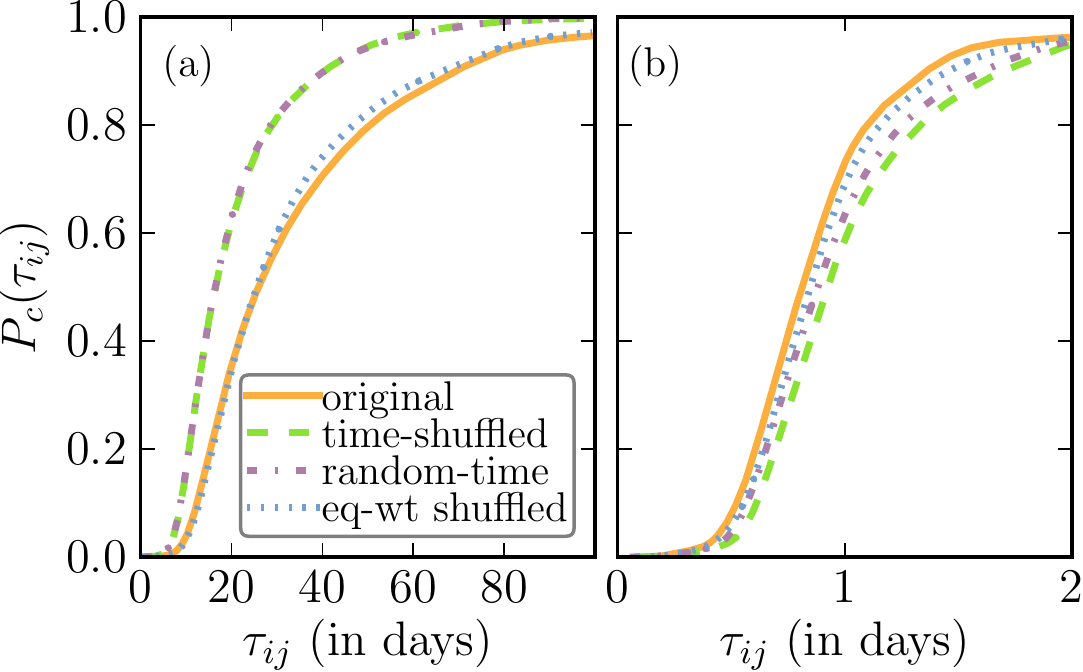}
\end{center}
\caption{Cumulative distribution of the temporal distances for the (a)
mobile phone call and (b) air transport network. The corresponding
distribution for the time-shuffled, random-time and equal-weight
link-sequence shuffled cases are also shown. It is seen that the distances
in the mobile phone call network are relatively long compared to the
time-shuffled and random references, whereas they are short in the air
transport transport network designed to transfer passengers in an optimal
way.}
\label{fig:tdistDistribution}
\end{figure}

It has been earlier seen for the full mobile communication network that the
burstiness of event sequences results in slower speed of SI
dynamics~\cite{Karsai10}.  This observation was based on simulated
spreading, averaged over a number of initial conditions. As such dynamics
follows shortest temporal paths, one would expect a similar effect on
average temporal path lengths in general. This is indeed the case.
Fig.~\ref{fig:tdistDistribution}(a) shows the cumulative probability
distribution (CDF) of temporal distances for the original sequence and null
models.  Clearly, distances are shorter for the time-shuffled and
random-time models where bursts are destroyed; the similarity of these
curves points out that the daily pattern plays a negligible role.  The
similarity of the CDF's for the original sequence and equal-weight
link-sequence shuffled model indicates that event-event correlations are
also fairly unimportant for temporal distances, in line
with~\cite{Karsai10}.

For the air transport network, the situation is strikingly different
[Fig.~\ref{fig:tdistDistribution}(b)].  The temporal distances in the
original case are lower than for any null model, indicating that overall,
the role of heterogeneities and correlations is to speed up dynamics in
this system. This is not surprising as the events of this transport network
are scheduled in an optimized way for the network to efficiently transport
passenger. Removing event-event correlations (the equal-weight
link-sequence shuffled model) is seen to slightly increase distances.  The
daily pattern is also seen to give rise to a minor increase in distances.

\begin{figure}
\begin{center}
  \includegraphics[width=0.85\linewidth]{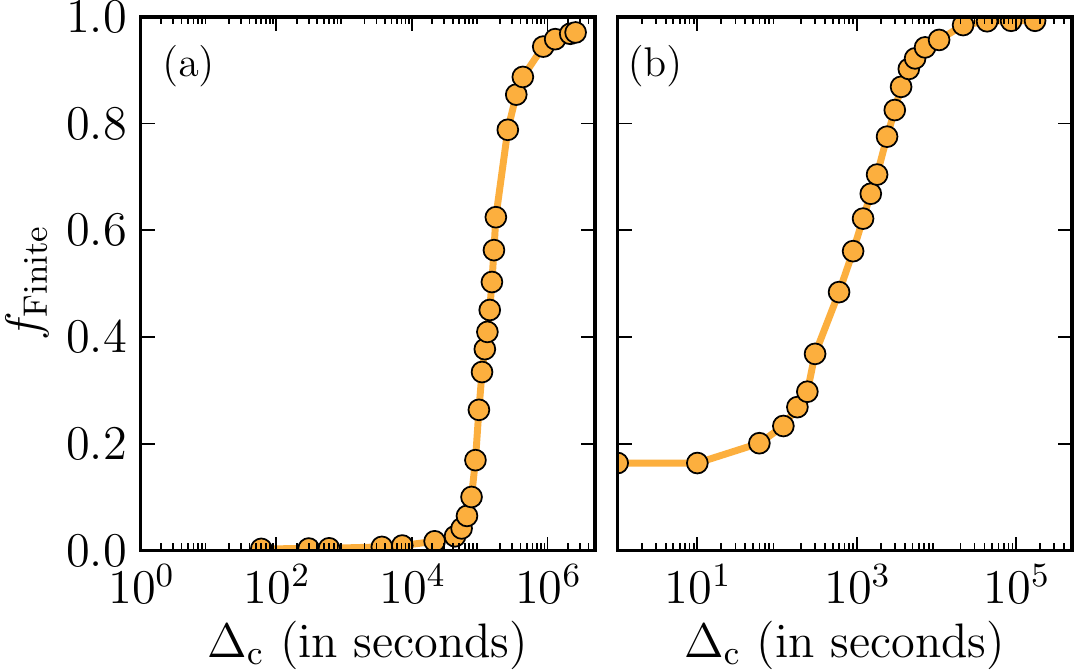}
\end{center}
\caption{Fraction of finite temporal paths as a function of
$\Delta_{\mathrm c}$ for the (a) mobile phone call and (b) air transport
network.}
\label{fig:fracFinite}
\end{figure}
\subsection{Temporal paths with waiting time cutoff}

So far, we have considered any sequence of events that follows temporal
ordering a valid path. Let us now introduce an additional criterion for the
existence of a path: the waiting time cutoff $\Delta_{\mathrm c}$,
indicating the maximum allowed time between two consecutive events on a
path. 
As an example, suppose there is an instantaneous event between nodes
$i$ and $j$ at time $t_1$, and another between $j$ and $k$ at time $t_2$.
These events then span the path $i\rightarrow j \rightarrow k$ only if the
time difference between the events $0<(t_2-t_1)\le\Delta_{\mathrm c}$. If
the events have an associated duration $\delta t$, the criterion becomes
$0<\left[t_2-\left(t_1+\delta t_1\right)\right]\le\Delta_{\mathrm c}$. If
spreading dynamics along such paths are considered, the cutoff makes such
dynamics SIR-like. In the SIR dynamics (Susceptible, Infectious, Recovered),
an infectious node remains infectious only for a limited period of time before
recovery and immunity to further infections. Hence, in such dynamics, 
for anything to be transmitted via a node, it has to be
transmitted quickly enough. In the context of mobile calls, the cutoff time means
that information is no longer passed on after a too long waiting time, \emph{i.e.}~it
becomes obsolete or uninteresting. Similarly, for the air transport network,
imposing a cutoff means that flights are not considered as connecting if the 
transit time is too long. Temporal paths constrained by the waiting
time cutoff are the paths along which such spreading or transport processes may take place.


The cutoff time $\Delta_{\mathrm c}$ restricts the number of allowed paths,
and we quantify this effect by calculating the overall fraction of node
pairs joined by finite temporal paths within the period of observation,
$f_{\mathrm{Finite}}$, also called the \emph{reachability
ratio}~\cite{Holme05a}, as a function of $\Delta_{\mathrm c}$. In the call
network, for low $\Delta_{\mathrm c}$, most nodes remain disconnected
[Fig.~\ref{fig:fracFinite}(a)]. However in the air transport network, even
when $\Delta_{\mathrm c} = $1 sec, $f_{\mathrm{finite}}=0.16$. This is
because of two factors: a large number of direct connections, and a large
number of simultaneous arrivals and departures at airports.  For both
networks, most pairs of nodes are eventually connected by temporal paths as
$\Delta_{\mathrm c}$ increases. For the call network, connectivity emerges
approximately when $\Delta_{\mathrm c} > 2$ days. Hence for any information
to percolate through this system, nodes should forward it for at least 2
days after its reception. 
This result is fairly surprising; such a long period severely constrains 
global information cascades. However, it is in line with earlier observations
that in simulations, structural and temporal features of call networks tend to limit
the flow of information~\cite{Onnela07,Karsai10}.
For the air transport network, most temporal
paths become finite when $\Delta_{\mathrm c} > 30$ minutes. This is
consistent with the minimum transit time required for catching a connecting
flight.


\begin{figure}
\begin{center}
  \includegraphics[width=0.85\linewidth]{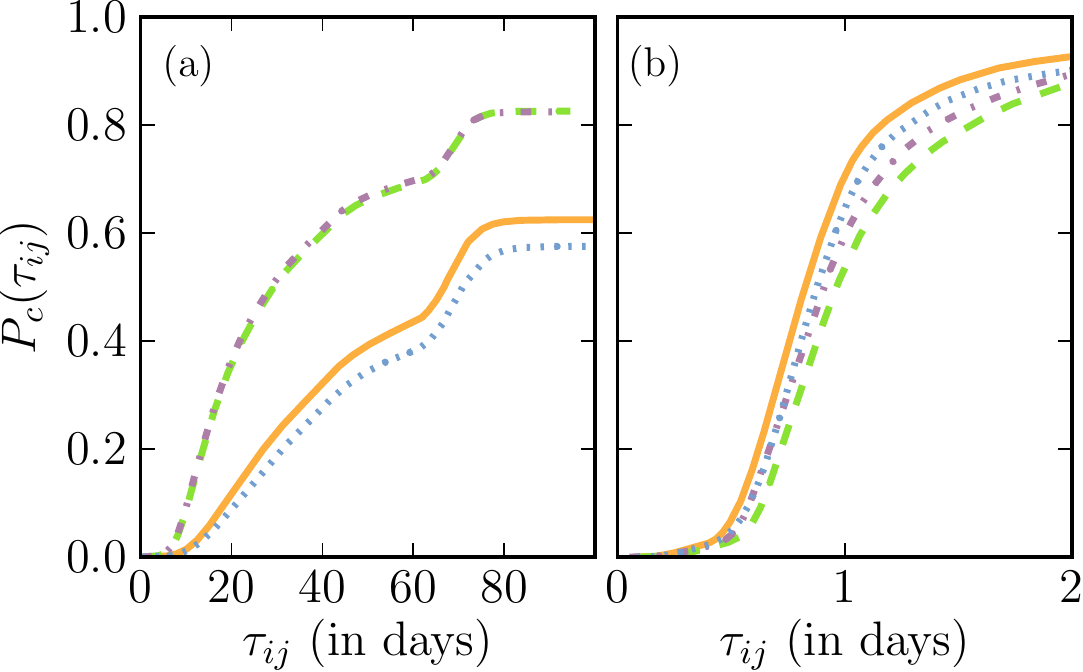}
\end{center}
\caption{Comparison of the cumulative probability distributions of the
temporal distances for the original and the randomized null models in the
(a) mobile phone call network with a cutoff $\Delta_c = 2$ days and (b) air
transport network with cutoffs $\Delta_c^{\mathrm{min}}$ = 30 min and
$\Delta_c$ = 5 hours. Line styles denote different null models, similarly to Fig.~4.}
\label{fig:tDelta}
\end{figure}

Let us next apply the null models and study temporal paths with cutoffs
$\Delta_{\mathrm c}$. For the call network [Fig.~\ref{fig:tDelta}~(a)], we
set $\Delta_{\mathrm c} =$ 2 days. The CDFs of temporal distances show that
only a fraction of finite temporal paths exists for all cases.  This
fraction is considerably larger for the time-shuffled and random-time null
models, as the bursty event sequences give rise to longer waiting times and
thus limit the number of existing paths. In addition, as above, the
temporal distances for these null models are on average lower than for the
original sequence, and hence also SIR-like dynamics is slowed down by
bursts. Further, event-event correlations, \emph{i.e.}, rapid chains of
calls $i\rightarrow j \rightarrow k$, make the paths somewhat faster, as
could be expected, since in the equal-weight link-sequence shuffled model
where such chains are destroyed the temporal distances are higher.  The
jump in the tail of the distribution is due to the finite 120-day period of
observation and a large number of pairs of nodes connected via two events
only, giving an average $t_{ij}\approx 60$ days.

For the air transport network, we apply an additional lower waiting time
cutoff to account for the time needed to catch a connecting flight, and
require the waiting times of between consecutive events to be between
$\Delta_{\mathrm c}^{\mathrm{min}}$ = 30 min and $\Delta_{\mathrm c}$ = 5
hrs. The order of the cumulative probability distributions of temporal
distances [Fig.~\ref{fig:tDelta}~(b)] for all the null models is similar to
the unconstrained case. Like for the call network,  event-event
correlations are seen to shorten temporal paths, as destroying them with
the equal-weight link-sequence shuffled model gives
rise to longer distances.

\subsection{Temporal Closeness Centrality}

So far, we have focused on the overall temporal distances that limit the
speed of any dynamics on temporal graphs. To conclude our investigation,
let us focus on the properties of individual nodes and their importance.
To measure of how quickly all other nodes can  be reached from a given
node, we define the \emph{temporal closeness centrality} as 
\begin{equation}
  C_{i}^{\mathrm T} = \frac{1}{N-1}\sum_j \frac{1}{\tau_{ij}},
\end{equation}
where $\tau_{ij}$ is the average temporal distance between $i$ and $j$ and
$N$ the number of nodes. A high value of $C_{i}^{\mathrm T}$ thus indicates
that other nodes can be quickly reached from $i$. This measure is a
generalization of the closeness centrality for static networks, defined as
the inverse of the average length of the shortest paths to all the other
nodes in the graph~\cite{Freeman79}: 
\begin{equation}
  C_{i}^{\mathrm S} = \frac{1}{N-1}\sum_j \frac{1}{d_{ij}}, 
\end{equation} 
where $d_{ij}$ is the static distance between the nodes $i$ and $j$. A high
value of $C_{i}^{\mathrm S}$ indicates that in the static network other
nodes can be reached in a few steps from $i$, whereas low value means that
other nodes are on average either unreachable or can only be reached via long
paths.\footnote{Note that for both cases, dynamic and static, we have chosen
to average over inverse distances rather than define the centrality as 
the inverse average distance. This choice has been made to better account for
disconnected pairs of nodes.} 

For comparing the static and temporal closeness centrality to topological
properties of nodes, we adopt the point of view
of spreading, where short distances to other nodes are likely to improve the efficiency
of the process, and central nodes are likely to be influential spreaders.
We study the dependence of the static and temporal closeness centrality of a node on two quantities:
node degree $k$ and its $k$-shell index, $k^s$. 
The node degree can 
be viewed as a first approximation of the importance of a node for spreading. 
However, it has recently been shown that in fact, the most efficient spreaders are located
within the core of the network, \emph{i.e.} have a high value of $k^s$~\cite{Kitsak10}.
The $k$-shell index of a node is an integer quantity, measuring its ``coreness".
To decompose the network into its $k^s$-shells, all nodes with degree $k=1$
are recursively removed until no more such nodes remain, and assigned to
the 1-shell.  Remaining higher-degree nodes are then recursively removed
for each value of $k$ and assigned to the corresponding shell, until no
more nodes remain.

\begin{figure}
\begin{center}
    \includegraphics[width=0.9\linewidth]{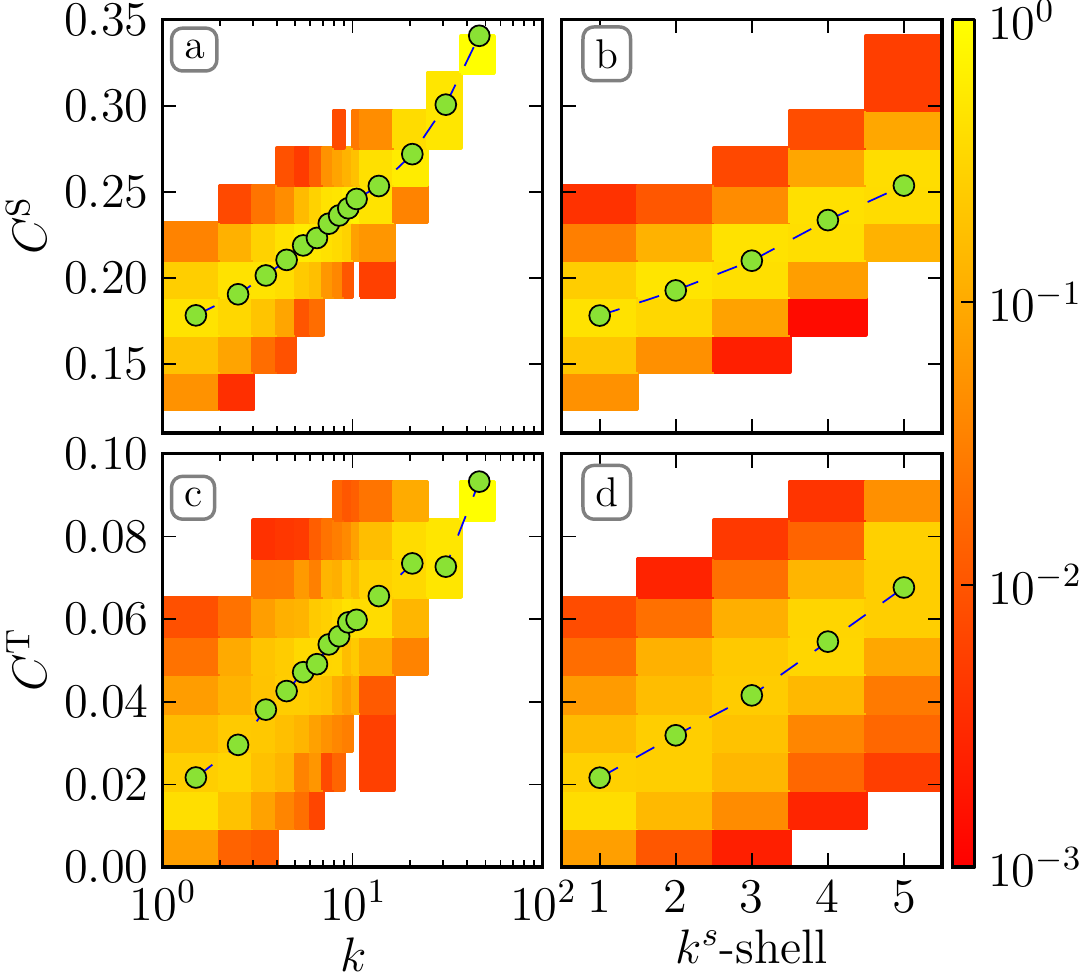}
\end{center}
\caption{Static and temporal closeness centrality ($C^{\mathrm S}$ and
$C^{\mathrm T}$) of the nodes against their (a,c) degree, $k$ and (b,d)
$k^s$-shell index in the mobile phone call network. Circles denote mean values,
while the shading represents
conditional probabilities $P(C^{\mathrm{S,T}}|k)$ and $P(C^{\mathrm{S,T}}|k^s)$.
}
\label{fig:centralityCity}
\end{figure}

The dependence of the static and temporal closeness centrality for the call network on both $k$ and $k^s$ is shown in Fig.~\ref{fig:centralityCity}.
Clearly, both quantities $C^{\mathrm S}$ and $C^{\mathrm T}$ increase with $k$ and $k^s$ on average.
 However, again there is a large spread around
the mean, and nodes with a high $k$ or $k$-shell index but a low static
or temporal closeness can be found. Measured with the linear Pearson
correlation coefficient, we find that the static $C^{\mathrm S}$ correlates with $k$
and $k^s$ with coefficient values of $C=0.80$ and $C=0.81$, respectively.
The correlation of the temporal $C^{\mathrm T}$ with $k$ and $k^s$ is
slightly weaker, with values of $C=0.69$ and $C=0.76$,
respectively. However, even these values are fairly high. 
Thus both the static and temporal closeness centralities are clearly
associated with high degrees and shell indices on average.


\begin{figure}
\begin{center}
    \includegraphics[width=0.9\linewidth]{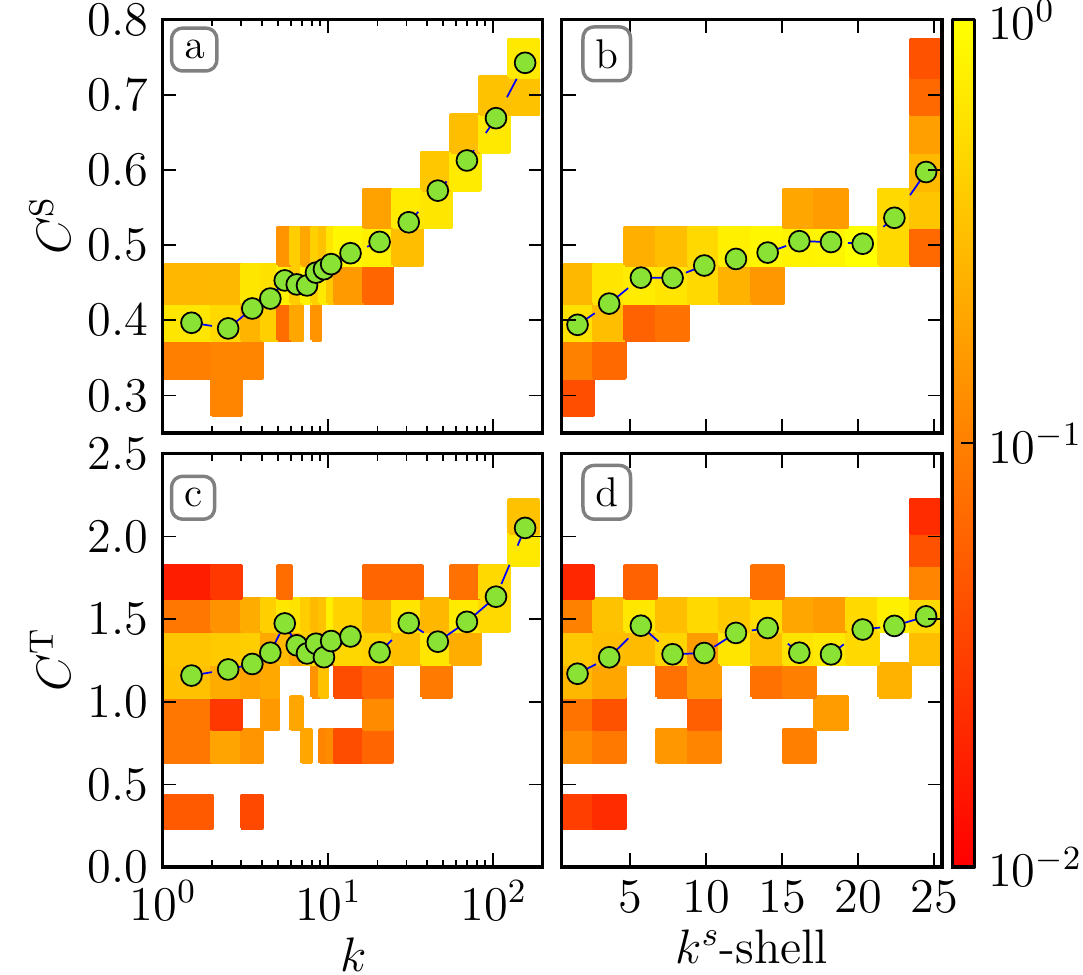}
\end{center}
\caption{
Static and temporal closeness centrality ($C^{\mathrm S}$ and
$C^{\mathrm T}$) of the nodes against their (a,c) degree, $k$ and (b,d)
$k^s$-shell index in the air transport network. Circles denote mean values,
while the shading represents
conditional probabilities  $P(C^{\mathrm{S,T}}|k)$ and $P(C^{\mathrm{S,T}}|k^s)$.
}
\label{fig:centralityAirline}
\end{figure}

For the air transport network, we find a
different result~(Fig.~\ref{fig:centralityAirline}). The static closeness
centrality $C^{\mathrm S}$ correlates strongly with
degree $k$ ($C=0.89$) and the $k^s$-shell index ($C=0.88$).
 However, the correlation between the temporal closeness centrality with
$k$ and $k^s$ is much weaker, with coefficient values $C=0.45$ and $C=0.46$,
respectively. 
The explanation for this observation is that the network is
geographically embedded, and temporal path lengths are heavily influenced by flight times, \emph{i.e.} the geographical distances between airports. 
Thus the nodes representing airports around the central regions of USA
should on average be connected to other airports by short temporal paths,
unless connected by a too low frequency of flights, whereas airports around 
the coastal areas should have lower temporal centralities.
Indeed, this is the case. When ranked according to $C^{\mathrm T}$,
the top three airports are ATL, Atlanta (rank=1, $k$=156,
$k^s$=25); ORD, Chicago (rank=2, $k$=133, $k^s$=25); DFW,
Dallas (rank=3, $k$=126, $k^s$=25).
These major airports have high values of $k$ and $k^s$, reducing the number of 
transfers needed to reach other airports, and are located away from the coast. 
There are also airports that have a high temporal centrality but low $k$ and 
$k^s$, typically located
in the central states of USA and also connected to other temporally
central nodes, e.g., CHA, Chattanooga (rank=8, $k$=5,
$k^s$=5);  MGM, Montgomery (rank=9, $k$=2, $k^s$=2); ACT, Waco
(rank=10, $k$=1, $k^s$=1). On the contrary, many interlinked coastal hubs
that score low in the temporal centrality ranking can be found in the highest $k^s$-shells:
e.g., IAD, Washington (rank=152,
$k$=64, $k^s$=25); MCO, Orlando (rank=79, $k$=69, $k^s$=25);
JFK, New York (rank=199, $k$=59, $k^s$=25).

\section{Conclusions and Discussion}

The properties of time-ordered temporal paths play a crucial role for any
dynamics taking place on temporal graphs, such as the flow of information
or resources or epidemic spreading. In essence, their maximum velocity  is
defined by the time it takes to complete such paths.  Building on a
definition of average temporal distance and its algorithmic implementation,
we have studied temporal paths in empirical networks. Although our results
show that temporal and static distances between nodes are correlated, in
general there is a wide spread.  Thus although nodes may be close in the
static network, the time it takes to reach one from another may be very
long, or vice versa, and in some cases, there is no temporal path at all.
Because of this, any spreading process may follow very different paths on
the temporal graph, and nodes that appear fairly insignificant from the
static network perspective may in fact rapidly transmit information or
disease around the network.  Second, as shown with null models, temporal
distances are affected by heterogeneities and correlations in the sequence
of events spanning the paths. In line with earlier observations, these were
seen to increase temporal distances for human communication networks --
however, for the air transport network, the optimized scheduling of flights
has the opposite effect.

Furthermore, we have also raised the issue of the finite observation period. 
For any measure to be applied on temporal graphs, the size and finiteness
of the time window are important issues. Here, we have taken care to 
define the average temporal distance such that unnecessary artifacts are avoided.
Yet, the application of this measure may yield results that are not useful if
the observation window is too short in relation to event frequency. 
On the other hand, if the observation window is too long, the system may 
undergo changes during the window (e.g. in terms of its node composition) that
make the results difficult to interpret.
Hence, for any analysis of temporal graphs, the observation
window issue is an important one, and further studies and methods for choosing a
proper window size are in our view called for.


Finally, it is worth stressing that the null models we apply retain both the
underlying network topology as well as the total numbers of events on each
of its edges; hence, depending on the temporal heterogeneities, the
dynamics of processes may differ a lot even when they take place on
networks that appear similar from the static perspective. This is
especially crucial for processes such as SIR spreading, where infection may
not be transmitted further if the waiting times between consecutive events
on temporal paths are too long.  Thus, in simulations and modeling of
processes such as epidemic spreading, information flow, and socio-dynamic
processes in general, the time-domain properties of the event sequences
that carry the interactions should be taken into account.

\begin{acknowledgments}
  Financial support from EU's 7th Framework Program's FET-Open to
  ICTeCollective project no. 238597 and by the Academy of Finland, the
  Finnish Center of Excellence program 2006-2011, project no. 129670,  are
  gratefully acknowledged.  We thank A.-L.~Barab\'asi for the mobile call
  data used in this research.
\end{acknowledgments}

\bibliography{temporal}

\appendix
\section{Algorithm for computing temporal distances}

\begin{figure}[bt!]
\begin{center}
  \includegraphics[width=0.9\linewidth]{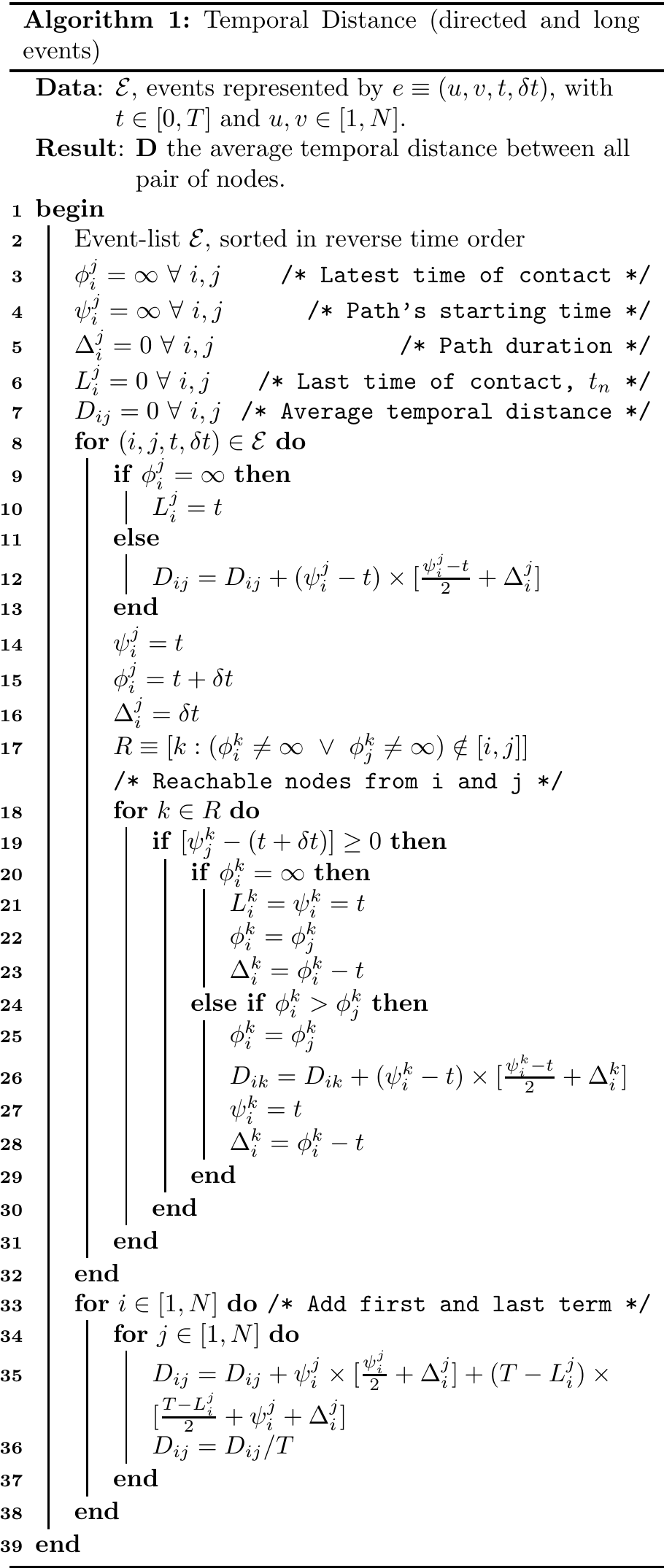}
\end{center}
\caption{Pseudo code for the temporal distance algorithm with directed and non-instantaneous events.}
\label{fig:algorithm}
\end{figure}

Here we present  the generalized temporal distance algorithm, where the
events are directed and/or have a duration to completion. The main flow of
the algorithm follows the instantaneous and undirected case, see Sect. II.
However, when the events are directed, for each event $(i,j,t)$ only the
vector clock of $i$ is compared element-wise with that of $j$, i.e.,
$\phi_{i}^{k}$ and $\phi_{j}^{k}$ $\forall k$. If
$\phi_{j}^{k}<\phi_{j}^{k}$, $\phi_{i}^{k}$ is replaced with
$\phi_{j}^{k}$, and we also set $\phi_{i}^{j}(t)=t$. The vector clock of
$j$ remains unchanged.  When the events also have an associated duration
$\delta t$,  we have to define an additional vector for each node,
$\pmb{\psi}_i$, which stores the last observed beginning times of temporal
paths from $i$ to all other nodes.  Like for ${\pmb{\phi_i}}$, the elements
of this vector are also set to $\infty$ in the beginning of a run. When
handling an event $(i,j,t,\delta t)$, the vector clock of $i$ is again
element-wise compared  to that of $j$, and if $\phi_{i}^k <\phi_{j}^k$ for
some $k$, it is checked if $\psi_{j}^k>t+\delta t$. If this condition
holds, the element $\phi_{i}^{k}$ is updated to $\phi_{j}^{k}$ and the
element $\psi_{i}^{k}=t$. 
One also sets $\phi_{i}^{j}(t)=t+\delta t$ and $\psi_{i}^{j}(t)=t$, since
$j$ can be reached from $i$ through an event starting at $t$ and finishing
at $t+\delta t$, and thus $\tau_{ij}(t)=\delta t$. A pseudo-code for the
algorithm is given below.

\end{document}